\documentclass{cosmo02}

 \usepackage{graphics,epsfig}

\begin{document}

\begin{frontmatter}



\title{Formation and Evolution of Galactic Black Holes}


\author{F. Combes}

\address{Observatoire de Paris, LERMA, 61 Av. de l'Observatoire,
F-75014, Paris, France}

\begin{abstract}
The main requirements for fueling an active galactic nucleus
and to form massive black holes are reviewed. 
Low-luminosity AGN 
can be fueled easily from the local star clusters,
near the nucleus, and the various stellar processes are 
described.
Above a certain luminosity (and therefore accretion rate) large-scale
gas flows from galactic scales are required. These can be driven
by gravity torques of non-axisymmetric perturbations, such as bars, 
spirals, galaxy interactions. Observational evidence that these
mechanisms are in action is found for high enough luminosities.
It is very frequent that starbursts are also triggered through
the same mechanisms, and the dense nuclear star clusters formed
provide fuel for the AGN over a longer time-scale.
 Secular internal evolution and more violent evolution through 
interactions and mergers contribute to grow both a massive black hole
and a bulge, and this could explain the observed proportionality 
relation between the mass of these two components.
\end{abstract}


\end{frontmatter}

\section{How to Fuel a Massive Black Hole}

From the observed AGN luminosities, and an
assumed conversion efficiency to transform the gravitational
energy into radiation, the order of magnitudes of the accretion
rates can be derived. Luminosities can be typically
of the order or higher than 10$^{46}$ erg/s.
If we assume a mass-to-energy conversion efficiency
$\epsilon \sim$ 10\%  (L = dM/dt c$^2$ $\epsilon$), then the
mass accretion rate dM/dt should be:

dM/dt $\sim$ 1.7 (0.1/$\epsilon$)  (L/10$^{46}$ erg/s)  M$_\odot$/yr

\noindent If the duty cycle of the AGN is of the order of
10$^8$ yr, then a total mass up to 2 10$^8$ M$_\odot$ should
be available.
It is a significant fraction of the gas content of a
typical galaxy, like the Milky Way!
The time-scale to drive such a large
mass to the center is likely to be larger than 1 Gyr.

For the mass to infall into the center, it must lose its
angular momentum. Could this be due to viscous torques?
In a geometrically thin accretion disk, one can consider the
gas subsonic viscosity, where the viscous stress is modelled
proportional ($\alpha$) to the internal pressure,
with a factor $\alpha < 1$.
This can only gather in 1 Gyr the gas within
4 $\alpha$ pc typically (e.g. Shlosman et al 1989, Phinney 1994).
 This shows that viscous torques will not couple the large-scale
galaxy to the nucleus, only the very nuclear regions could
play a role through viscous torques.

\subsection{Stars as the AGN Fuel}

The stars themselves could provide gas to the nucleus, through
their mass loss, if there is a local stellar cluster, dense and compact
enough (core radius R$_c$ of less than a pc, core mass
M$_{core}$ of the order of 10$^{8}$ M$_\odot$).
However, the mass loss rate derived from normal stellar evolution gives
only 10$^{-11}$ M$_\odot$/yr/M$_\odot$,
orders of magnitude below the required rate of a few  M$_\odot$ /yr.
 The contribution will be significant, only
if a massive stellar cluster (4 10$^9$
M$_\odot$) has just formed through a starburst
(Norman \& Scoville 1988).
A coeval cluster can liberate 10$^9$ M$_\odot$ on 10$^8$ yr,
since mainly massive stars evolve together in the beginning.
 Thus the existence of a starburst in the first place solves
also the problem of the AGN fueling, as in the
symbiotic model of Williams et al (1999). The angular momentum
problem is now passed on to the starburst fueling, and could
be solved only through large-scale dynamical processes.

It is also possible that stars themselves are
directly swallowed by the black hole, the various processes
that have been studied are:
\begin{itemize}
\item Bloated stars, a phenomenon that makes mass loss more efficient
 (Edwards 1980, Alexander \& Netzer 1994, 1997),
\item Tidal disruption of stars (Hills 1975, Frank \& Rees 1976),
\item Star Collisions (Spitzer \& Saslaw 1966, Colgate 1967, Courvoisier 
et al 1996, Rauch 1999).
\end{itemize}

\begin{figure}[t]
\rotatebox{-90}{\includegraphics[width=.90\textwidth]{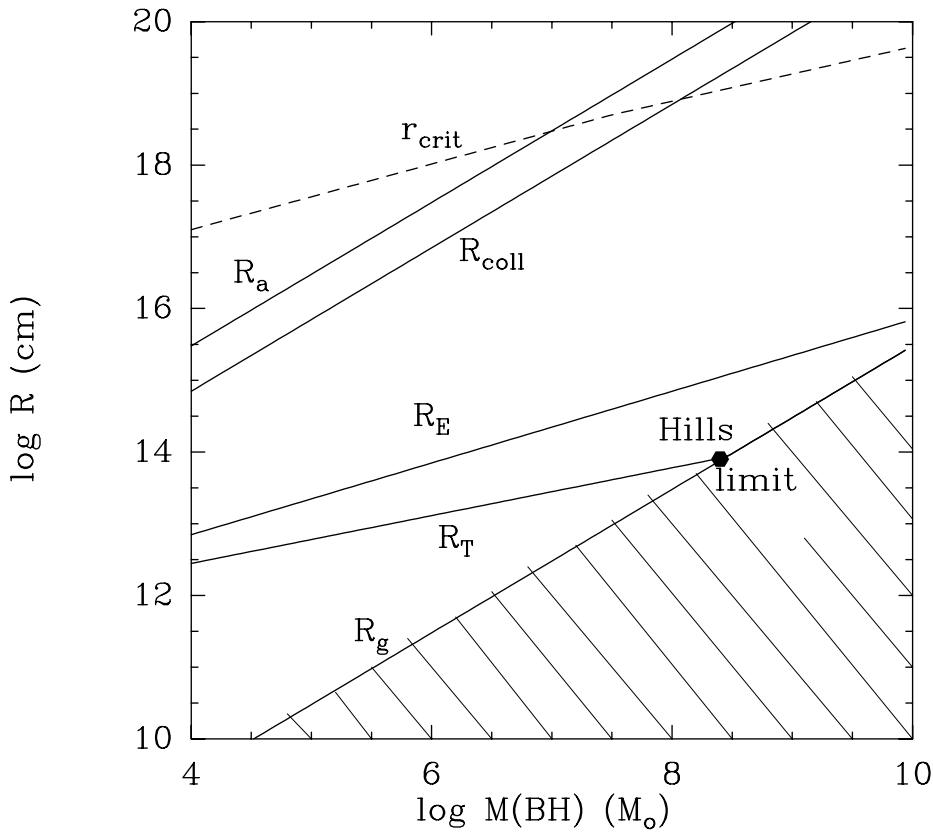}}
\caption{ Characteristic radii, corresponding to the various physical
phenomena, as a function of black hole mass M$_\bullet$: from top to bottom,
R$_a$, the accretion radius, 
under which the BH dominates the dynamics;
R$_{coll}$, the collision radius,
  under which stellar collisions are disruptive, i.e. the free-fall velocity
 around the black hole (GM$_\bullet$/r)$^{1/2}$ is equal 
 to the escape speed v$_*$
of a typical individual star (GM$_*$/r$_*$)$^{1/2}$;
R$_E$, the Eddington radius,  
under which a star receives more light than its Eddington luminosity;
R$_T$, the tidal radius, 
under which a star is disrupted by the tidal forces of the BH;
R$_g$, the gravitational radius. 
Loss-cone effects are important inside the critical radius r$_{crit}$.}
\label{char}
\end{figure}

The characteristic radii associated
to the prevalence of these processes are displayed in figure
\ref{char}.
As the black hole horizon grows faster with  M$_\bullet$ than the 
tidal radius, there is a limit, when M$\sim$ 3 10$^8$ M$_\odot$, above 
which the star disruption occurs inside the black hole, 
and there is no gaseous release or AGN activity 
(but the black hole might grow even more rapidly). This is the
Hills limit (Hills 1975).

\begin{figure}
\rotatebox{-90}{\includegraphics[width=.45\textwidth]{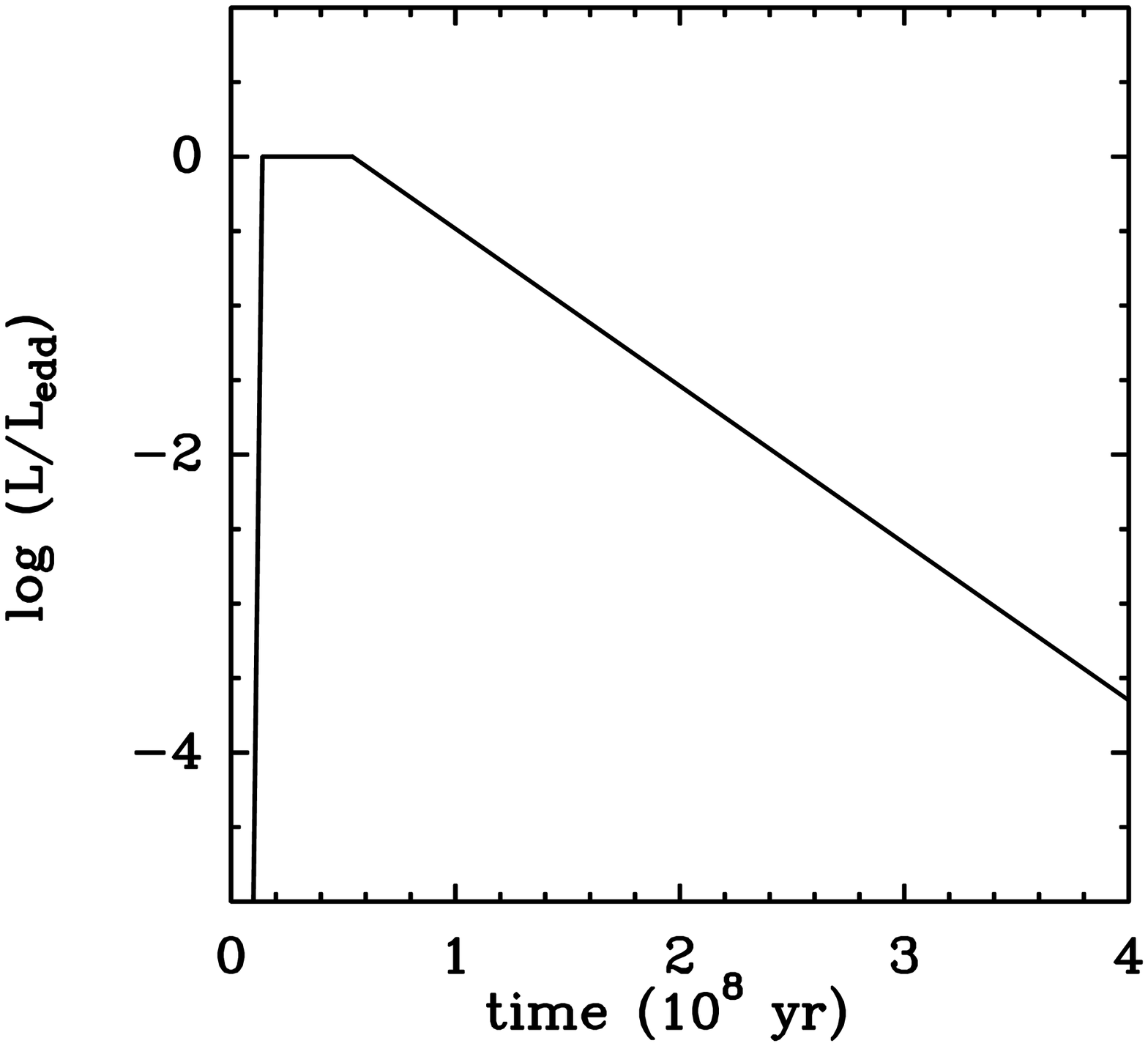}}
\rotatebox{-90}{\includegraphics[width=.45\textwidth]{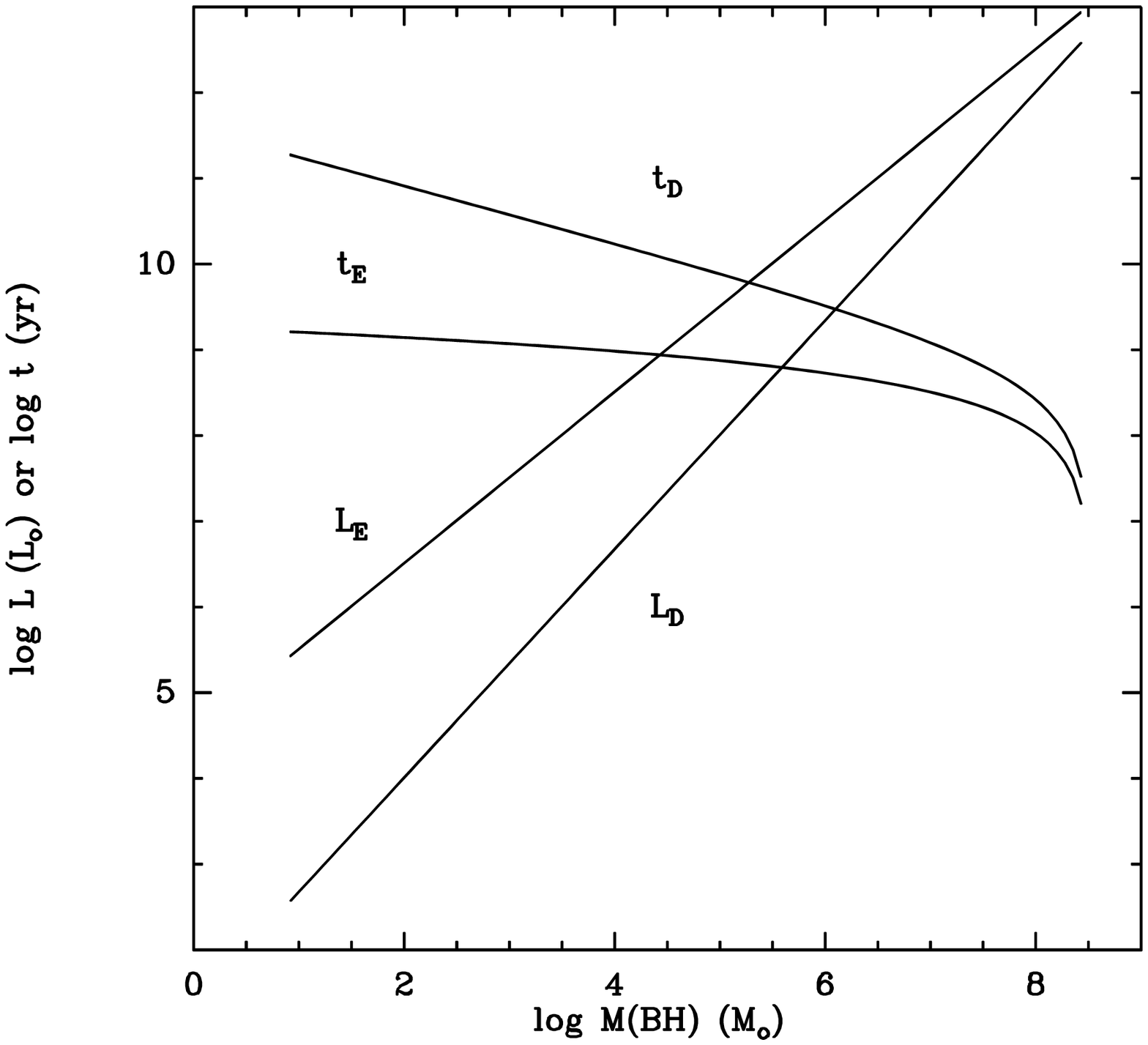}}
\caption{ {\bf (left)} 
 A schematic view of the activity phase of a typical
quasar: during the growth phase, the AGN does not radiate
efficiently, although the fueling rate is larger than Eddington.
After its luminous phase of  $\sim$ 4 10$^7$ yr, the fuel
is exhausted and the AGN fades away.
{\bf (right)} 
Growth of a supermassive black hole in two simple models:
accretion at Eddington luminosity (time-scale t$_E$ and corresponding
luminosity L$_E$, as a function of black hole mass M$_{BH}$), and
when accretion is limited by diffusion (t$_D$ and L$_D$)
(from Hills, 1975).
}
\label{hills}
\end{figure}

\subsection{Growth of Black Holes}

There is now a consensus to recognize that AGN derive their power
from supermassive black holes, but their formation
history, their demography, their activity time-scales
are still debated. The two extreme hypotheses have been
explored: either only a small percentage of
galaxies become quasars, and they are continuously fueled,
and active over Gyrs, or a massive black hole exists in
almost every galaxy, but they have active periods of only
a few 10$^7$ yr.  In the first hypothesis, there should exist
black holes with masses 100 times higher than the maximum
observed today, and accretion rates much lower than
the Eddington rate, and this is not supported
(e.g. Cavaliere \& Padovani 1989).
Models with a duty cycle of 4 10$^7$ yr are favored,
and many galaxies today should host a starving
black hole (Haehnelt \& Rees 1993). Figure \ref{hills}a
schematizes the period of activity and growth of
a typical massive black hole.  It is also possible
that further gas accretion (during a galaxy interaction
for instance) triggers a new activity phase for
the black hole.

If we assume that the availability of the fuel around the black hole is 
not a problem, then the black hole can
at maximum accrete mass at the Eddington rate, and radiate at Eddington 
luminosity (above which the radiation
pressure prevents the material to fall in). A typical growth rate for the 
black hole is then given by 
the time required to reach the critical mass
M$_c$ where R$_T$ = R$_g$, above which stars are
swallowed by the black hole without any gas radiation
(M$_c$ = 3 10$^8$ M$_\odot$).
The Eddington luminosity is:
L$_E$ = 3.2 10$^4$ (M/M$_\odot$) L$_\odot$.
For a mass M$_c$, the maximum is 10$^{13}$ L$_\odot$ (close to the
peak luminosity of QSOs). Then the corresponding accretion rate,
assuming an efficiency of $\epsilon$ =10-20\%  is
dM/dt$_E$ = 1.1 10$^{-8}$ (M/M$_\odot$) M$_\odot$/yr.
The growth rate of the black hole in this regime t$_E$ is then exponential;
it takes only 1.6 10$^9$ yr to
grow from a stellar black hole of 10 M$_\odot$ to M$_c$:

t$_E$= 9.3 10$^7$ ln(M$_c$/M) yr

\noindent Note that this very simple scheme would lead to  a
maximum at z=2.8 of the number of quasars.
This maximum rate, however, is not realistic, since
the black hole quickly gets short of fuel, as the neighbouring
stars  (in particular at low angular momentum)
are depleted.  Then it is necessary to consider a
growth limited by stellar density $\rho_s$:

DM/dt = $\rho_s \sigma$ V

\noindent where $\sigma$ is the accretion cross-section, and
V the typical stellar velocity. The corresponding
time-scale to grow from M to M$_c$ is

\noindent t$_D$ = 1.7 10$^{15}$ yr ($\rho_s$/M$_\odot$pc$^{-3}$)$^{-1}$
M/M$_\odot^{-1/3}$
(1 - M/M$_c^{1/3}$) $<$V$^2>^{1/2}$ (km/s)
\noindent Typically in galaxy nuclei,
 $\rho_s$ = 10$^7$ M$_\odot$/pc$^3$,  $<$V$^2>^{1/2}$  = 225 km/s.
A black hole could grow up to M$_c$ in a Hubble time, and
the luminosity at the end could be of the order of
10$^{46}$ erg/s (see figure \ref{hills}b).

In more details, the first stars to disappear, being swallowed by the black 
hole, are those with the least angular momentum, with a small pericenter. The
 only mechanism to replenish the stellar density near the BH is the two-body 
relaxation, with a time-scale $t_R$. 
The relevant two-body relaxation time
t$_R$ is dependent on the number of bodies N in the system,
as  t$_R$/t$_c$ = N/logN, where  t$_c$ is the crossing time= r$_c$/V.
For a galactic center, with a volumic density of stars of
10$^7$M$_\odot$/pc$^3$, this relaxation time is
 3 10$^8$ yr.

Since the angular momentum diffuse faster than the energy, the low anglular 
momentum stars will re-appear faster, this is the loss-cone effect, that increases
 the accretion rate by a factor $(1-e^2)^{-1}$, where $e$ is the exentricity of the orbits.
This is significant inside a critical radius r$_{crit}$, where the loss-cone angle
becomes larger than the diffusion angle $\theta_D \sim (t_{dyn}/t_R)^{1/2}$.
This critical radius is also plotted in figure \ref{char}.

\subsection{Formation of a Cusp of Stars around the BH}
\begin{figure}
\rotatebox{0}{\includegraphics[width=.40\textwidth]{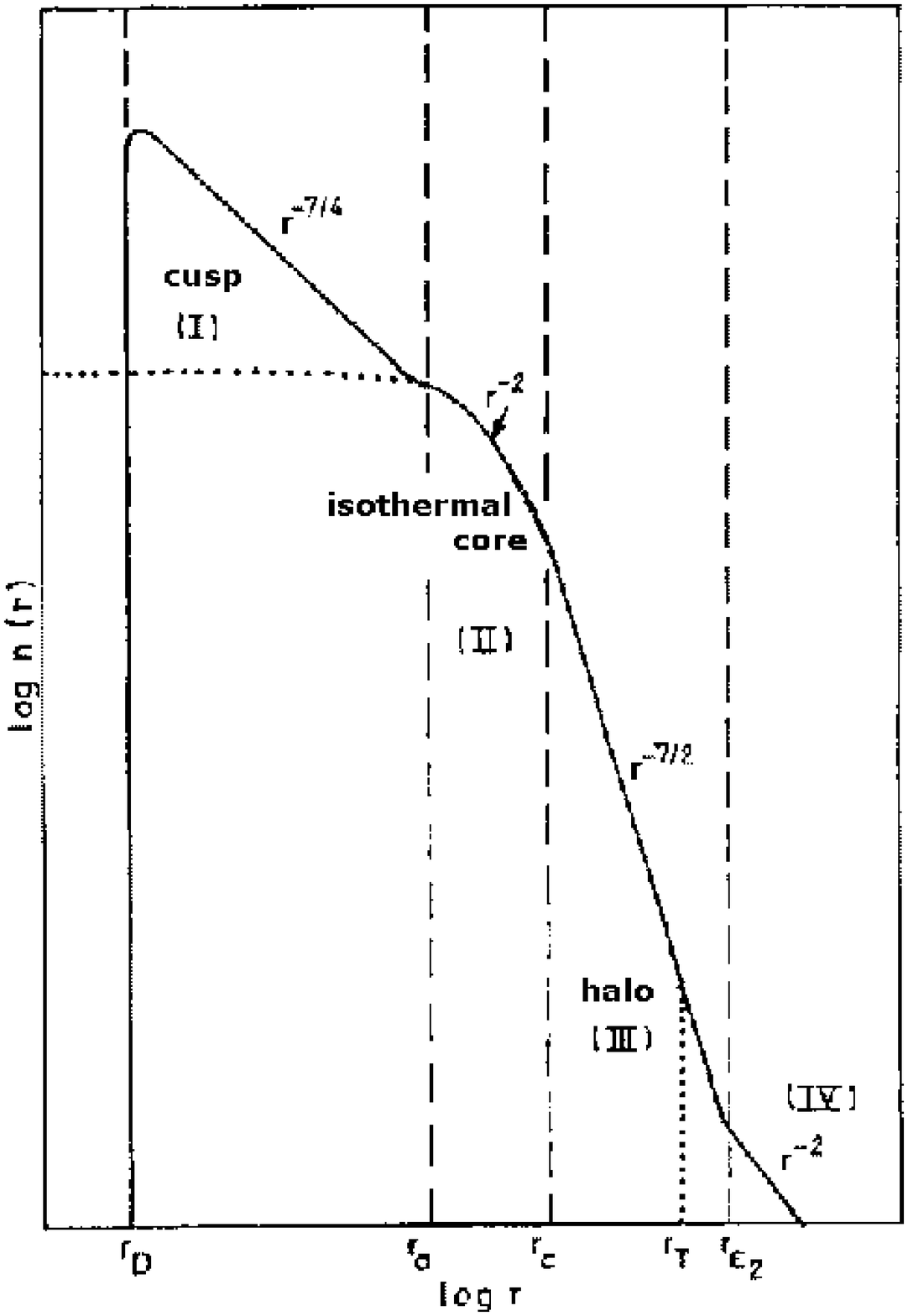}}
\rotatebox{-0}{\includegraphics[width=.60\textwidth]{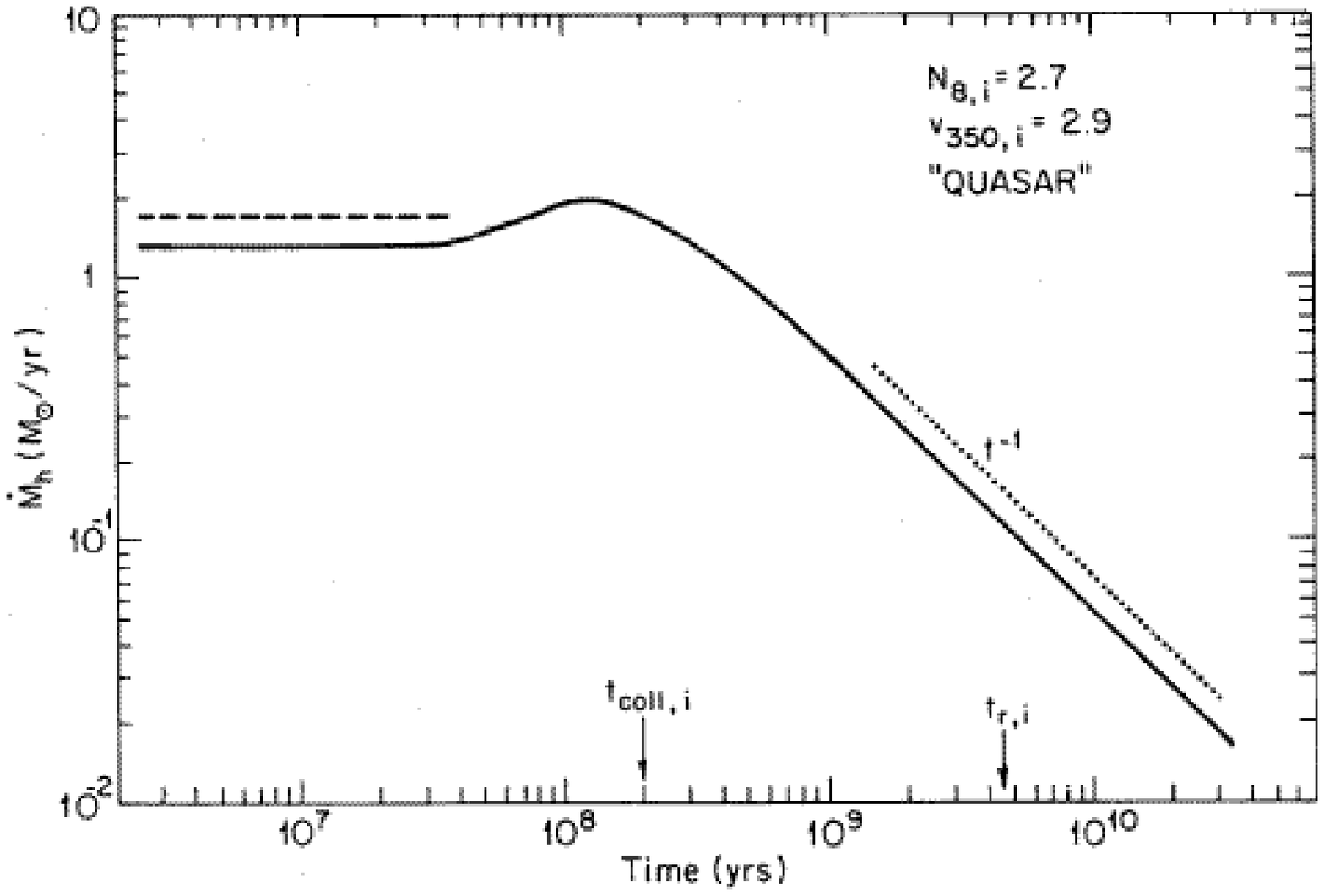}}
\caption{ {\bf (right)} 
 A schematic representation of the radial profile of stellar density
in a spheroid with a central massive black hole. (I) the cusp, 
with a slope of r$^{-7/4}$, (II) an isothermal core (in r$^{-2}$),
(III) a halo (in r$^{-7/2}$), and (IV) an isothermal tail (in r$^{-2}$).
{\bf (left)} the black hole growth rate, as a function of time,
according to a model where the number od stars in the core is
initially 2.7 10$^8$, and the velocity dispersion in the center
2.9 x 350 = 1015 km/s, parameters fitted to a ``quasar'' environment;
the dash-curve is the initial rate predicted analytically
(from Duncan \& Shapiro, 1983).}
\label{duncan}
\end{figure}

From a numerical resolution of the time-dependent Boltzmann equation,
with the relevant diffusion coefficients, it can be shown that around
a black hole at the center of a globular cluster, the stellar density should
be of a power-law shape, with a slope of 7/4, cf fig \ref{duncan}a
 (Bahcall \& Wolf 1976).

The distribution of stars around a black hole can be described,
according to the distance to the center:
\begin{itemize}
\item first for the stars not bound to the black hole,
at r $>$ R$_a$, their velocity distribution is
Maxwellian, and their density profile has the
isothermal power law in r$^{-2}$. There are also
unbound stars inside R$_a$, but with a density
in  r$^{-1/2}$. This allows to compute the penetration
rate of these unbound stars in the tidal or collision radius,
to estimate the accretion rate. With a core stellar mass of
M$_{core}$ = 10$^7$- 3 10$^8$ M$_\odot$, a density 10$^7$ pc$^{-3}$,
the accretion rate is, by tidal disruption:

 dM/dt$_{tide}$ = 1 M$_\odot$/yr M$_8^{4/3}$

\noindent and by stellar collisions:

dM/dt$_{coll}$ = 0.1 M$_\odot$/yr M$_8^3$

\item the orbits bound to the black hole  r$<$ R$_a$: due to the cusp,
their density is in  r$^{-7/4}$, there is an excess of stars
inside R$_{coll}$, that favors stellar collisions.
\end{itemize}

However, detailed numerical 
show that the stellar cluster cannot fuel the black hole
indefinitely (Duncan \& Shapiro 1983).
The growth rate of the black hole and its luminosity
decreases as 1/time (cf fig \ref{duncan}b).
The loss-cone theory and the simulations are in
agreement:
the accretion rate due to tidal disruptions  is
M$_{core}$/t$_R$, typically of 10$^{-2}$ M$_\odot$/yr, with a
 maximum lower than 1 M$_\odot$/yr; this cannot explain
the luminosity of QSOs. QSOs might be explained only when stellar
collisions are included, the corresponding accretion
rate is typically a hundred times higher.

Triaxial deviations from spherical symmetry of
only 5\%  (due to a bar or binary black hole)
can repopulate the loss-cone, increasing tidal
disruption to QSOs levels.
However, t$_{coll} <$ t$_R$, and collisions may destroy
the cusp (Norman \& Silk 1983).

Stellar collisions help to refill the loss-cone,
although they flatten the stellar cusp (Rauch 1999).
The collisions rate is comparable to the diffusion
rate, that refill the central core.

In summary,
for low density nuclei, stellar evolution and
tidal disruption is the main mechanism
to bring matter to the black hole, and
for high density nuclei, stellar collisions
dominate the gas fueling.
The evolution of the stellar density through these processes
is then opposite, and accentuates the differences:

-- for n $<$ 10$^7$/pc$^3$ the core then expands, due to
heating that results from the settling of a small
population of stars into orbits tightly bound to
the black hole;

-- for  n$>$ 10$^7$/pc$^3$, the core shrinks due to the
removal of kinetic energy by collisions.
To give an order of magnitude, the nuclear density
in our own Galaxy is estimated at
10$^8$ M$_\odot$/pc$^3$ (Eckart et al 1993).

These mechanisms produce differing power-law
slopes in the resulting stellar density cusp
surrounding the black hole, -7/4 and -1/2
for low- and high-density nuclei, respectively
(Murphy et al 1991, Rauch 1999).
In simulations however (Rauch 1999), collisions tend
to produce a flat core, instead of r$^{-1/2}$ law in
Fokker-PLank studies which imply isotropy
(and are unable to treat sparse regions).

The cusps observed in nearby galaxies are consistent with the
hypothesis of adiabatical black hole growth
in homogeneous isothermal core (e.g. Young 1980)
 and with initial conditions for the cores following 
the scaling relations of the
fundamental plane (van der Marel 1999).  One must also take
into account that the merging of black holes, and the dynamical
effects of binary black-holes, flatten the cusp slopes
(Nakano \& Makino, 1999; Milosavljevic \& Merritt, 2001).

\subsection{Conclusion for AGN Fueling by Stars}

Stars are sufficient to fuel some low luminosity AGN: there is 
first mass loss from a nuclear cluster, which provides only low 
accretion rates, then tidal disruption of stars themselves, which 
depletes the center, replenished through two-body relaxation. At high 
densities stellar collisions also replenish the central density, and the
 AGN can reach higher luminosities. According to the central density, the
 power-law slopes of the density cusp in the center are different (flatter
 for higher densities). However, the accretion rates reached through fueling
 by stars only are not sufficient to account for the most luminous AGN and
 quasars. Large-scale processes to transfer angular momentum of the 
interstellar gas over a large disk radius are then required.

\section{Radial Mass Flows through Bars and Spirals}

Gravity torques created in galactic disks by
non-axisymmetrical perturbations, due to gravitational instabilities, can 
efficiently transfer the angular momentum of the interstellar gas, and help 
to abundantly fuel the nucleus. Numerical simulations, supported by observations,
 have established in the recent years that bars and spirals are
density waves permanently renewed in galaxy disks. 

The sign of the gravity torques, and consequent radial mass flows depend on the
 position of resonances, between the motions of the particles and the density wave
 pattern; the main features of the orbits in a barred potential and the associated
 resonances are now briefly recalled.

\subsection{Resonances}

The stellar orbits in a nearly axisymmetric
potential $\Phi$ are at first order circular with an angular
velocity $\Omega^2 = {1\over r}{d\Phi\over dr}$. In linearizing the
potential in the neighborhood of a circular orbit, the motion of any
particle can be expressed in first order by an epicyclic oscillation,
of frequency $\kappa$,
$$
\kappa^2 = r {d\Omega^2\over dr} +4\Omega^2
$$
The general orbit is therefore the combination of a circle and an
epicycle, or a rosette, since there is no rational relation between
the two periods.

\begin{figure}
{\centering \leavevmode
\epsfxsize=.45\textwidth \rotatebox{-90}{\epsfbox{combesf-f4a.ps}}\hfil
\epsfxsize=.45\textwidth \rotatebox{-90}{\epsfbox{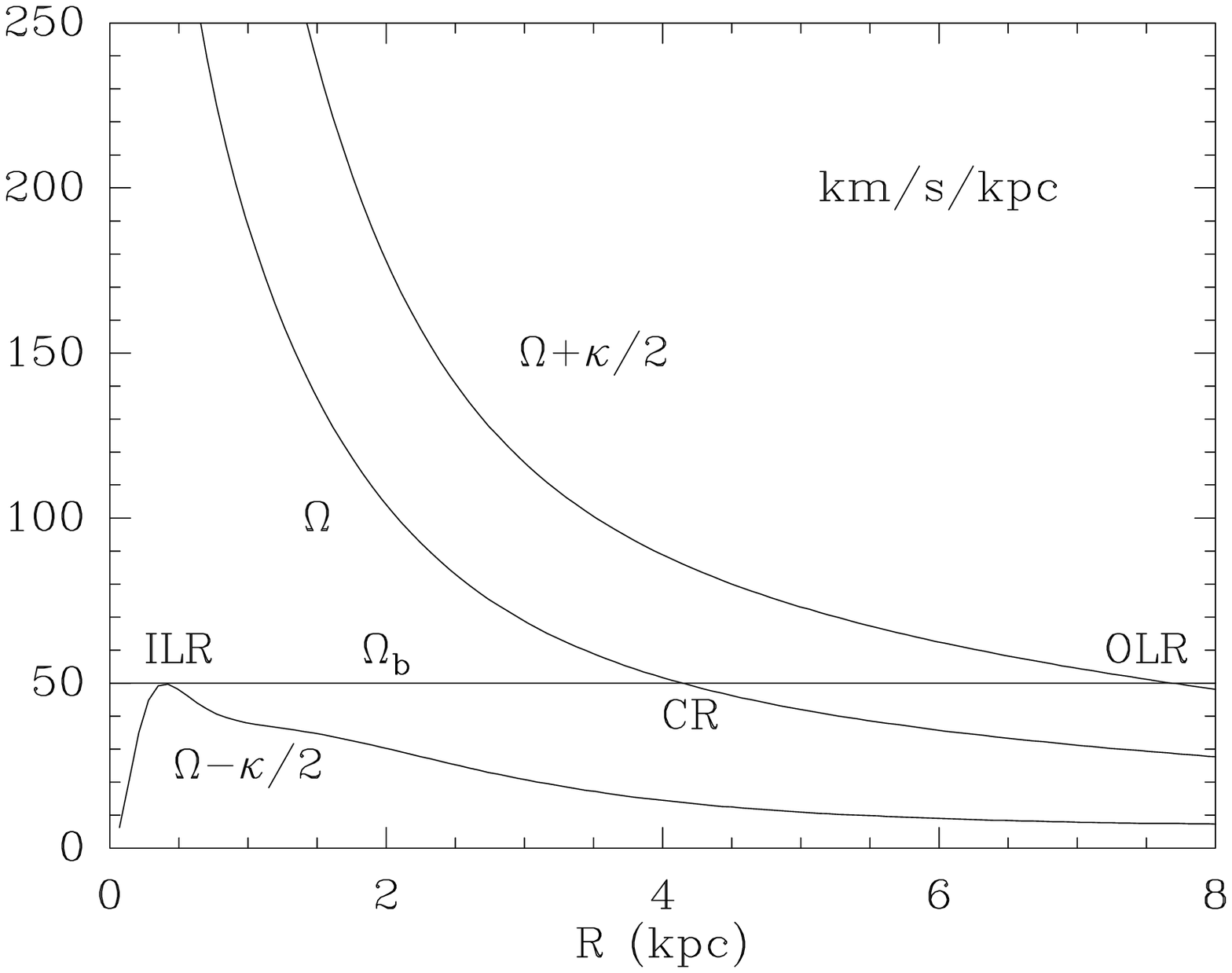}}}
\caption{ {\bf left}: Various types of resonant orbits in a galaxy.
 At the ILR, the orbit is closed and elongated, and
is direct in  the rotating frame. At Corotation (CR), the orbit
makes no turn, only an epicycle; at OLR, the orbit is closed, elongated
and retrograde in  the rotating frame. In between, the orbits
are rosettes that do not close.
 {\bf right}: Frequencies $\Omega$, $\Omega - \kappa/2$ and $\Omega + \kappa/2$
in a galaxy disk. The bar pattern speed  $\Omega_b$ is indicated, defining the locations
of the Linblad resonances. }
\label{resa}
\end{figure}

The bar creates a bisymmetric gravitational potential, with a predominant
Fourier component $m=2$, which rotates in the galaxy with the pattern speed
$\Omega_b$. There is a region in the plane where the pattern speed
is equal to the frequency of rotation $\Omega$, and where particles
do not make any revolution in the rotating frame. This is the resonance
of corotation (cf fig \ref{resa}).

Periodic orbits in the bar rotating frame are orbits that close on
themselves after one or more turns. Periodic orbits are the building
blocks which determine the stellar distribution function, since they
define families of trapped orbits around them. Trapped orbits are
non-periodic, but oscillate about one periodic orbit, with a similar
shape. The various families are (Contopoulos \& Grosbol 1989):
\begin{itemize}
\item the $x_1$ family of periodic orbits is the main family supporting
the bar. Orbits are elongated parallel to the bar, within corotation.
They can look like simple ellipses, and with energy increasing, they
can form a cusp, and even two loops at the extremities.
\item  the $x_2$ family exists only between the two inner Lindblad
resonances (ILR), when they exist. They are more round, and elongated
perpendicular to the bar. Even when there exist
two ILRs in the axisymmetric sense, the existence of the $x_2$ family
is not certain. When the bar is strong enough, the $x_2$ orbits
disappear. The bar strength necessary to eliminate the $x_2$ family
depends on the pattern speed $\Omega_b$: the lower this speed, the
stronger the bar must be.
\item  Outside corotation, the 2/1 orbits (which
close after one turn and two epicycles) are run in the retrograde sense
in the rotating frame; they are perpendicular to the bar inside
the outer Lindblad resonance (OLR), and parallel to the bar slightly
outside.
\end{itemize}

\begin{figure}
\begin{center}
\rotatebox{-90}{\includegraphics[width=.5\textwidth]{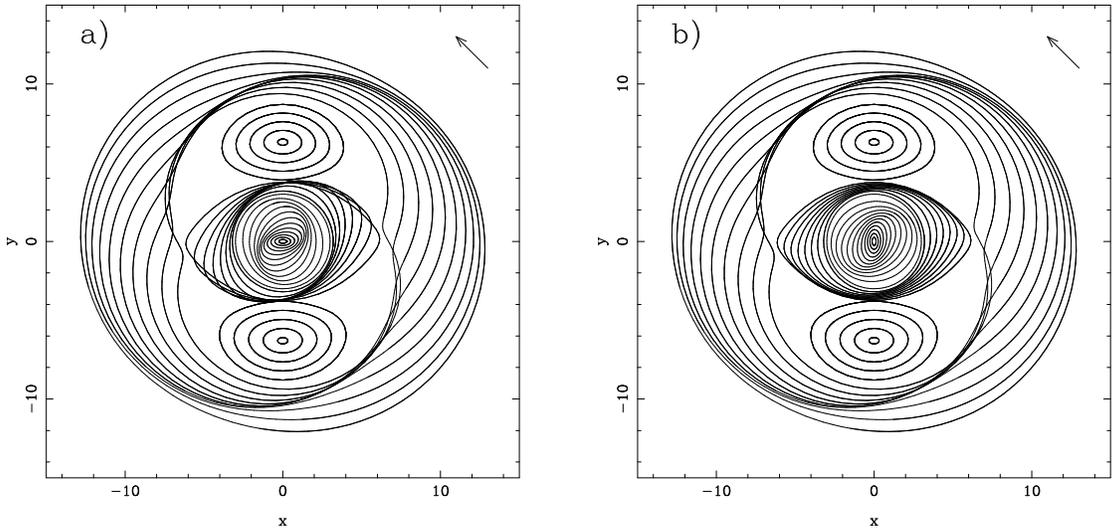}}
\end{center}
\caption{  Behaviour of the gas flow
in a barred galaxy ($\cos 2 \theta$ potential,
oriented horizontally). The gas tends to follow the periodic
orbits whose orientation rotates by $\pi/2$ at each resonance.
 Due to dissipation, the gas orientation change gradually.
The sense of winding can be derived from the variation of the precession rate
$\Omega-\kappa/2$ as a function of radius:
{\bf a)} Without a central mass concentration, $\Omega-\kappa/2$
is increasing with radius in the center: the gas winds up
in a leading spiral inside the ILR ring; {\bf b)} with a central mass
concentration, it is the reverse and the gas follows a trailing spiral structure,
inside ILR.}
\label{fuel2}
\end{figure}

In summary,
the orientation of the periodic orbits rotates by 90$^\circ$ at each
resonance crossing, and they are successively parallel and perpendicular
to the bar.  The gas will first tend to follow
these orbits, but the streamlines of gas cannot cross. Since periodic
orbits do cross, gas clouds can encounter enhanced collisions, such
that their orbits are changed. Instead of experiencing sudden
90$^\circ$ turns, their orbits will smoothly and gradually turn,
following the schematic diagram of kinematic waves, first
drawn by Kalnajs (1973), and illustrated in fig \ref{fuel2}.

This interpretation predicts that the arms will be more wound when there
exist more resonances; there will be a winding over 180$^\circ$ with only
CR and OLR, with the gas aligned with the bar until corotation. When there
exists 2 ILRs, the gas response can be perpendicular to the stellar bar.

\subsection{Angular Momentum Transfer}
\label{amt}

To minimize its total energy, a galaxy tends to concentrate
its mass towards the center, and to transfer its angular momentum
outwards (Lynden-Bell \& Kalnajs 1972). It is
the role of the spiral structure to transport angular momentum
from the center to the outer parts, and only trailing waves can do it.
This transfer, mediated by non-axisymmetric instabilities,
is the motor of secular evolution of galaxies, and of the formation of
bars and resonant rings.

\begin{figure}
\begin{center}
\rotatebox{-90}{\includegraphics[width=.50\textwidth]{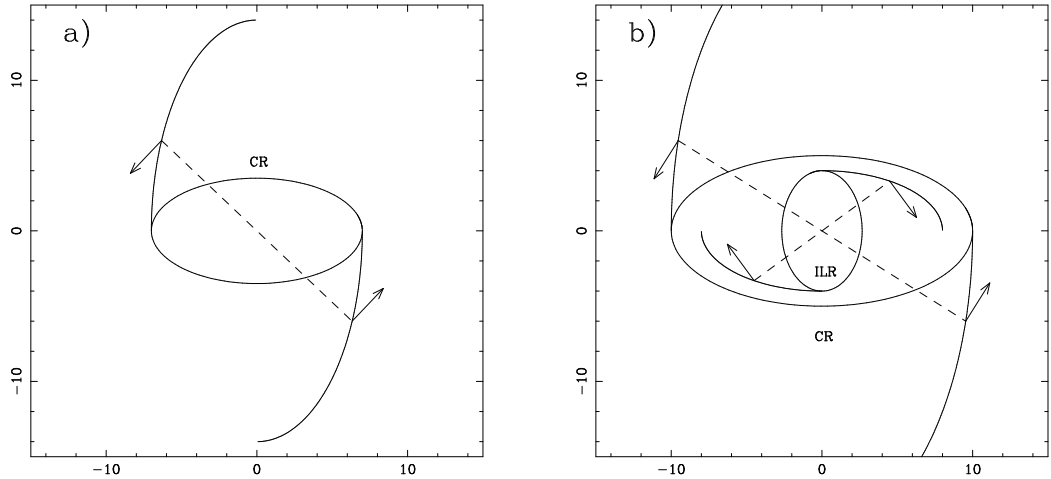}}
\end{center}
\caption{Schematic representation of the gravity torques exerted by
the bar on the gaseous spiral: {\bf a)} between CR and OLR, the gas acquires
angular momentum and is driven outwards; {\bf b)} between CR and ILR, the gas
loses angular momentum (cf Combes 1988). }
\label{torq}
\end{figure}

The angular momentum transfer is due to the torques
exerted by the bar on the matter forming spiral arms.
There is a phase shift between the density and the potential
wells, resulting in torques schematized in fig \ref{torq}.
The gas is much more responsive to these torques, since
they form the spirals in a barred galaxy.
 The torque changes sign at each resonance, where the
spiral turns by 90$^\circ$.  Between the ILR and corotation, the torque
is negative, while between CR and OLR, the torque is positive.  These
torques tend to depopulate the corotation region, and to  accumulate
gas towards the Lindblad resonances, in the shape of rings. Indeed,
these rings then are aligned with the symmetry axis of the bar, and
no net torque is acting on them. Numerical simulations of colliding gas
clouds in a barred potential show that rings form in a few dynamical
times, i.e. in a few Gyrs for the outer ring at OLR (Schwarz 1981),
or in $\sim$ 10$^8$ yr for nuclear rings at ILR (Combes \& Gerin 1985).

\subsection{Fueling Nuclear Activity}
\label{barfuel}

The main issue to fuel the nucleus is to solve
the transfer of angular momentum problem.
 Torques due to the bar are very efficient, but
gas can be stalled in a nuclear ring at ILR.
Other mechanisms can then be invoked:
viscous torques, or
dynamical friction of giant clouds (GMC) against stars.
The viscosity is in general completely unefficient
over the galactic disk, but the corresponding time-scale
is decreasing with decreasing radius. Unfortunately,
in the center of galaxies, the rotation is almost
rigid, the shear is considerably reduced, and so are the
viscous torques.
The time-scale for dynamical friction becomes
competitive below r=100pc from the center (about
10$^7$ (r/100pc)$^2$ yr for a GMC of 10$^7$ M$_\odot$).
 For the intermediate scales, a new mechanism is required.

Note that if there is a supermassive black hole in the
nucleus, it is easier to bring the gas to the center.
Indeed, the presence of a large mass can change the
behaviour of the precessing rate of orbits $\Omega-\kappa/2$:
instead of increasing with radius  inside ILR (as in fig \ref{resa}),
it will decrease.

Due to cloud
collisions, the gas clouds lose energy, and their galactocentric
distance shrinks. Since it tends to follow the periodic orbits,
the gas streams in elliptical trajectories
at lower and lower radii, with their major axes leading more and more
the periodic orbit, since the precession rate (estimated by $\Omega
-\kappa/2$ in the axisymmetric limit, for orbits near
ILR, and by $\Omega+\kappa/2$ near OLR) increases with decreasing radii
in most of the disk (fig \ref{fuel2}). This regular shift forces the gas into a trailing
spiral structure, from which the sense of the gravity torques can be
easily derived. Inside corotation, the torques are negative, and the
gas is driven inwards towards the inner Lindblad resonance (ILR).
Inside ILR, and from the center, the precessing rate is increasing with
radius, so that the gas pattern due to collisions will be a leading
spiral, instead of a trailing one (see Figure~\ref{fuel2}a). The gravity
torques are positive, which also contributes to the accumulation of gas
at the ILR ring. This situation is only inverted in the case of a
central mass concentration (for instance a black hole), for which the precession
rate $\Omega -\kappa/2$ is monotonically increasing towards infinity
with decreasing radii. Only then, the gravity torques will pull the gas
towards the very center, and ``fuel'' the nucleus.

The problem reduces to forming the black hole in the
first place. We show next that the accumulation of matter towards the center
can produce a decoupling of a second bar inside the primary bar. This
nuclear bar, and possibly other ones nested inside like russian dolls,
can take over the action of gravity torques to drive the gas to the
nucleus, as first proposed by Shlosman et al. (1989).

\subsection{Decoupling of a Nuclear Disk and/or Bar}

The bar torques drive progressively more mass towards the center.
 This matter, gaseous at the beginning, forms stars, and
gradually contributes to the formation of the bulge, since
stars are elevated above the disk plane, through vertical
resonances with the bar (e.g. Combes et al. 1990, Raha et al.
1991). When the mass accumulation is large enough,
then the precessing rate $\Omega-\kappa/2$ curve
is increasing strongly while the radius decreases, and
this implies the formation of two inner Lindblad resonances.
In between the two ILRs, the periodic orbits are
perpendicular to the bar ($x_2$ orbits), and the bar
loses its main supporters. The weakening of the
primary bar, and the fact that the frequencies of
the matter are considerably different now between the
inner and outer disk,
forces the decoupling of a nuclear disk, or nuclear bar from
the large-scale bar (primary bar).

\begin{figure}
{\centering \leavevmode
\epsfxsize=.60\textwidth \rotatebox{-90}{\epsfbox{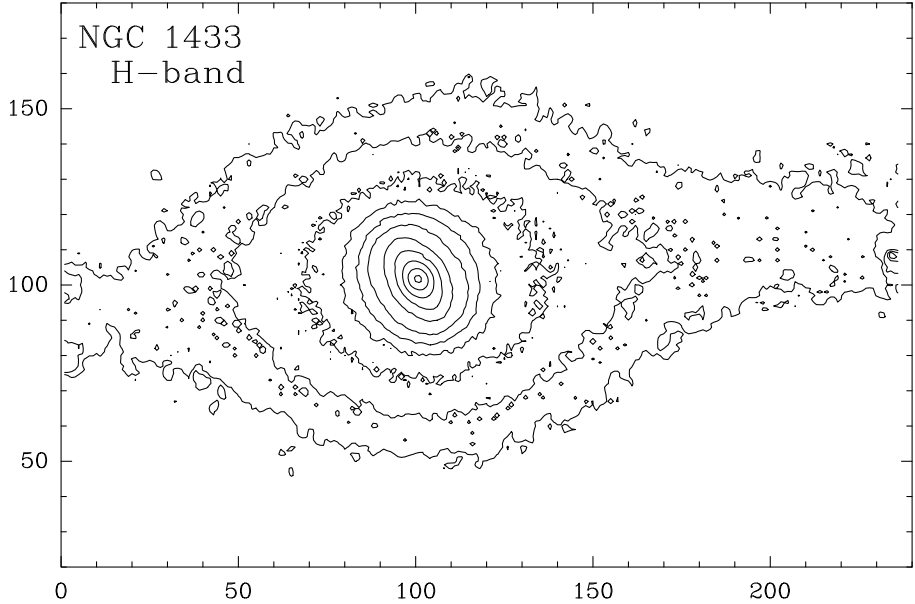}} }
\caption{ Example of a nuclear bar seen in the NIR H band:
the Seyfert 2 barred spiral galaxy NGC 1433. The pixel size is 0.52'',
from Jungwiert et al. (1997)}
\label{jung}
\end{figure}

Nuclear disks are frequently observed, at many wavelengths:
optical or NIR with the HST (e.g. Barth et al 1995, Regan \& Mulchaey 
1999), see also fig \ref{jung}),
or in CO molecules with millimeter
interferometers (Ishizuki et al 1990, Sakamoto et al 1999).
A recent survey in the Virgo cluster (Rubin et al
1997) reveals that about 20\% of the 80 spirals observed
possess a decoupled nuclear disk. 
The stellar nuclear disk is kinematically cold,
which suggest a recent formation from gas (Emsellem et al. 2001).
The percentage of double-barred
galaxies is observed around 20\% 
(Wozniak et al. 1995, Jungwiert et al 1997, Laine et al 2002).

\subsection{Bar Destruction and Renewal}
\label{bardest}

The inflow of matter in the center can destroy the bar.
It is sufficient that 5\% of the mass of the disk has sunk inside
the inner Lindblad resonance
(Hasan \& Norman 1990, Pfenniger \& Norman 1990, Hasan et al 1993).
But this depends on the mass distribution, on the size
of the central concentration; a point mass like a black hole is
 more efficient (may be 2\% is sufficient).
 The destruction is due to the mass re-organisation,
that perturbs all the orbital structure: the $x_1$ orbits sustaining
the bar for instance are shifted outwards.
Near the center, the central mass axisymmetrizes
the potential. Then there is a chaotic region,
and outside a regular one again.
 When a central mass concentration exists initially, in
N-body simulations, a bar still forms, but dissolves more
quickly. It is also possible that after a bar has dissolved,
another one forms, after sufficient gas accretion to
generate new gravitational instabilities: the location
of the resonances will not be the same
(Bournaud \& Combes 2002).

There might be several bar episodes in a
galaxy disk, with secondary bars transiently 
prolonging the action of the primary bars. The
succession of bars follow a self-regulated process, based on gravitational 
instability, and disc cooling through the gas dissipation. When a strong bar 
is developping, the gas inside corotation is driven inwards, and produce a 
central concentration that progressively tends to weaken or destroy the bar. 
At the same time, gas from the outer parts is held outside, since gravity 
torques are positive there. When the bar disappears, gas can then infall from
 the outer parts, and settles in the disk, that becomes unstable again to bar
 formation, etc..

When gas is driven inwards, it can fuel a starburst there, and then the central
 black hole. The vertical resonances help also to grow the bulge, in parallel
 to the black hole (e.g. Combes 2000).

\section{Are there Correlations between Bars and AGN?}

\begin{figure}
{\centering \leavevmode
\epsfxsize=.75\textwidth \rotatebox{-90}{\epsfbox{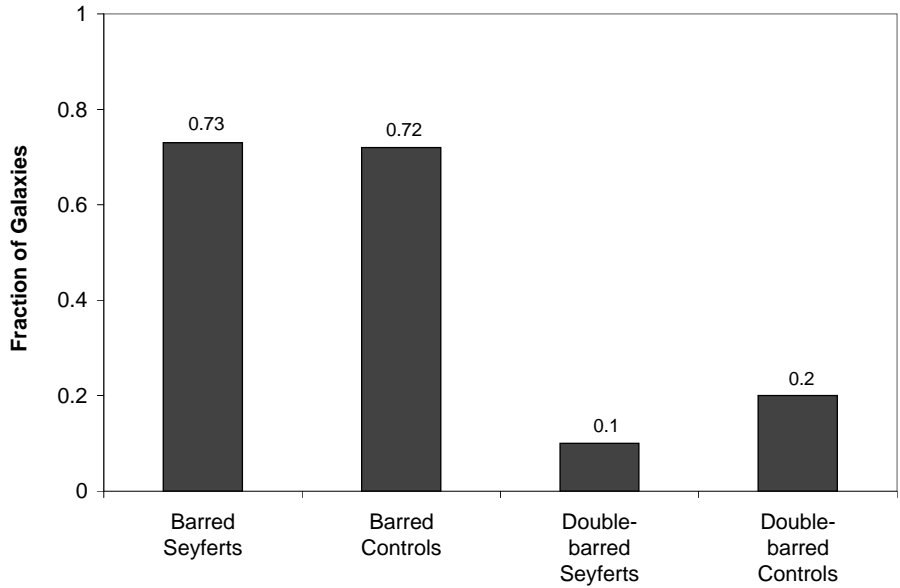}}}
\caption{ Fraction of barred and double-barred systems in the Seyfert and
control galaxy samples. The percentage of barred galaxies is comparable
in the Seyfert and control samples. A slightly higher percentage of the normal
galaxies have double bars (from Mulchaey \& Regan 1997).}
\label{mul97}
\end{figure}

There have been several observational works revealing
a correlation between nuclear activity and bars (Dahari 1984,
Simkin et al 1980) or between activity and distorted morphologies (Moles et al
 1995). But the correlation is
weak and depends on the definition of the samples,
their completion and other subtle effects. Near-infrared
images have often revealed bars in galaxies previously
classified unbarred, certainly due to gas and dust effects.
However, Seyfert galaxies observed in NIR do not statistically
have more bars nor more interactions than a control sample,
cf fig \ref{mul97}
(McLeod \& Rieke 1995, Mulchaey \& Regan 1997).

It is however evident from observations
that bars are efficient to produce radial gas flows:
barred galaxies have more H$_2$ gas concentration
inside their central 500 pc than un-barred galaxies
(cf Sakamoto et al 1999). Also, the radial flows
level out abundance gradients in barred galaxies
(Martin \& Roy 1994).

Peletier et al (1999) and Knapen et al (2000) have recently re-visited the 
question, and find that Seyfert galaxies are more barred than non-Seyfert, 
a result significant at at 2.5 $\sigma$.
They measure
the bar strength by the observed axial ratio in the images.
In Seyferts, the fraction of strong bars is lower than in
the control sample (Shlosman et al. 2000).
Although a surprising result a priori, this is not unexpected,
if bars are believed to be destroyed by central mass
concentrations (cf section \ref{bardest}).
Regan \& Mulchaey (1999) have studied 12
Seyfert galaxies with HST-NICMOS. Out of the 12, only 3 have
nuclear bars but a majority show nuclear spirals. However their
criterium for nuclear bars is that there exist leading dust lanes
along this nuclear bar. This is not a required characteristic, since
these secondary bars in general are not expected to have ILRs themselves.
Circumnuclear dust and nuclear spirals are frequently found, a striking result 
from HST color images 
(Martini \& Pogge 1999, Pogge \& Martini 2002).

The frequency of nuclear bars happen to be the same
(of the order of 20-30\% in active and non-active galaxies (Regan \& Mulchaey 
1999, Laine et al 2002).

If there is no correlation, this could be due to the fact that a strong bar 
is not sufficient to trigger nuclear activity, there should exist also a 
massive black hole in the center, and this favors the early-type galaxies. 
Indeed, there is a correlation between the mass of the bulge and that of the 
BH, and also, the fueling mechanism requires a central mass concentration to 
be efficient (see previous section).
Observationally, there is a good correlation between AGN and morphological types: 
AGN tend to lie in early-type galaxies (Terlevich et al 1987, Moles et al 1995, 
Ho et al 1997).
The lack of correlation could also be related to the variable gas content near the center. 
However, there does not appear to be more circumnuclear dust in Seyfert galaxies
 with respect to non active galaxies (Pogge \& Martini 2002).

A morphological study of the 891 galaxies in the Extended 12 $\mu$m
Galaxy Sample (E12GS) has confirmed that
Seyfert galaxies and Liners have the same percentage of bars as normal spirals,
contrary to HII/Starburst galaxies that have more bars
(Hunt \& Malkan 1999). However, active galaxies show rings significantly
more often than normal galaxies or starbursts.
The Liners have more inner rings (by a factor 1.5), while Seyferts have more
outer rings (by a factor 3-4) than normal galaxies.
This might be due to the different time-scales for bar and ring formation:
bars form relatively quickly, in a few 10$^8$ yr; they can drive
matter to the central regions and trigger a starburst there, in the
same time-scale. Outer rings form then, also under the gravity torques
of the bar, but in the dynamical times of the outer regions, i.e.
 a few 10$^9$ yr. Since Seyferts are correlated with them, they would be
associated to delayed consequences of the starburst, or of the bar,
which by this time begins to dissolve.

The percentage of AGN in the E12GS sample is 30\%. As for interactions,
25\% of the Seyferts are "peculiar" (disturbed), while 45\% of the
HII/Starbursts are. There is also a correlation between AGN and
morphological types along the Hubble sequence.
Seyferts tend to lie in early-types (Terlevich et al
1987, Moles et al 1995). This has to be related to the existence
on inner Linblad resonances in early-types, favoring the fueling
of the nucleus (e.g.  section \ref{barfuel}, Combes \& Elmegreen 1993).
In summary, there might be some evidence of the role of
large-scale dynamics on AGN fueling, but it is in general weak,
except for the most powerful AGN.
A good correlation between bars/interactions
and AGN is not expected, from several arguments:

-- there must be already a massive black hole in the nucleus,
and this might be the case only for massive-bulge galaxies (not all
barred galaxies)

-- again a large central mass concentration (bulge) is necessary to
produce an ILR and drive the gas inwards  (early-types)

-- other parameters, like geometrical parameters, control
the fueling efficiency of interaction

-- time-scales are not fitted: the AGN fueling is
postponed after the interaction/bar episode

-- there are other mechanisms to fuel AGN such
that a dense nuclear cluster

\section{Galaxy Interactions and Merging}

Galaxy interactions create also strong non-axisymmetries in galaxy disks, 
and are an efficient way to transfer angular momentum. They trigger strong 
bars in the interacting galaxy disks (e.g. Barnes \& Hernquist 1992) and drive
 the gas inwards through the same mechanisms as described before (section \ref{amt}).
Again, the fueling is dependent on the morphological type of the galaxy, since
 the stability of disks is essentially dependent on the bulge-to-disc ratio 
(Mihos \& Hernquist 1994, 1996). Numerical simulations have shown that the 
star formation rate is considerably enhanced in galaxy interactions, because
 of the gas concentration (the star formation is locally proportional to the 
gas density), and to the orbit crowding (schocks favoring the dissipation of 
gas). When the time-scale of radial gas inflow is smaller than the star-formation
 feedback (through supernovae, stellar winds, etc..), huge startbursts may occur.

There is indeed a very good correlation between 
high luminosity starbursts and galaxy interactions and
mergers (e.g. Sanders \& Mirabel 1996). For the
most luminous objects, a significant fraction of their 
luminosity is coming from nuclear activity, and
this fraction is increasing with infrared luminosity,
from 4 to 45\% (Kim et al 1998).
QSOs appear more than usual to interact with 
companions (Hutchings \& Neff 1992; Hutchings 
\& Morris 1995).
For AGN of low luminosity, external triggering appears less
necessary, since only 0.01 M$_\odot$/yr is required for
a Seyfert like NGC 1068 for example, during 10$^8$ yr.
However, once interactions drive gas to the
nucleus, some activity must be revived.
Time-scales may be the reason why the actions
are not simultaneous.
Large-scale gas has to be driven at very small
scales in the center, and the whole process requires
several intermediate steps.

Even for the good starburst/interactions
correlations, there are exceptions for the low-
luminosity samples.
Interacting galaxies selected optically (not IRAS galaxies)
are often not enhanced in star formation (Bushouse
1986, Lawrence et al 1989).
Only the obviously merging galaxies, like the
Toomre (1977) sample, are truly a starbursting
class; it is difficult to reveal a progression along a
possible evolutionary sequence (Heckman 1990).
There are too many determining parameters:
geometry, distance, gas content, etc...

A complication comes from the time-scales involved:
the starburst phase is short, of the order of
a few 10$^8$yr, similar only to the end of the merging
phase. It is more the presence of morphological
distortions than the presence of nearby
companions that is correlated with activity.
More than 50\% of ULIRGS possess multiple nuclei
(Carico et al 1990, Graham et al 1990).

Are Seyfert galaxies preferentially interacting?
According to Dahari (1984), 15\%  have close companions,
compared to 3\%  in the control sample.
But the Seyferts with or without companions have
the same H$\alpha$ or radio power (Dahari 1985),
although they may be more infrared bright with companions
(Dahari \& DeRobertis 1988, McKenty 1989).
According to Keel et al (1985),  there are 5\% of Seyferts
in control sample, and 25\%  in the close pairs of Arp Atlas.
But it is possible that the Arp Atlas galaxies suffer from
selection effects. Bushouse (1986) and deRobertis et al 1996) on the contrary finds a
deficiency of Seyferts in interacting galaxies.
A recent study by Schmitt (2001) finds 
no statistical difference in companion frequency for the various
activity types: Seyfert, Liners, transition, HII.
The claim that
Seyfert 2 have a larger number of companions than Seyfert 1
(Laurikainen \& Salo 1995, Dultzin-Hacyan et al 1999) was not 
confirmed by all studies (Schmitt et al 2001); this
could be an artifact, some Seyfert 2 have an UV excess due to a starburst 
 (Cid Fernandes et al 1998), and companions enhance the star formation rate.

\begin{figure}
{\centering \leavevmode
\epsfxsize=.40\textwidth \rotatebox{-0}{\epsfbox{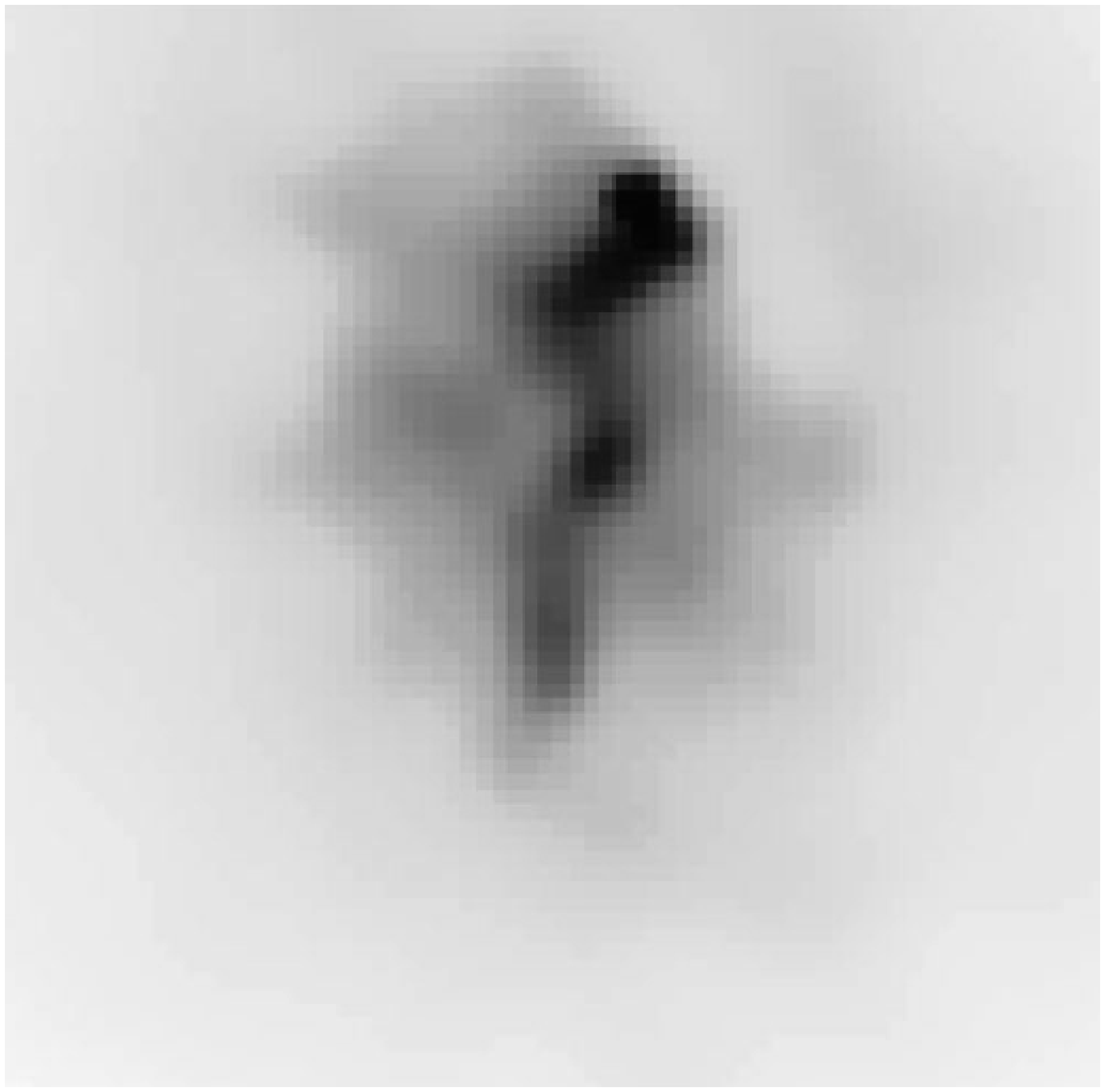}}\hfil
\epsfxsize=.40\textwidth \rotatebox{-0}{\epsfbox{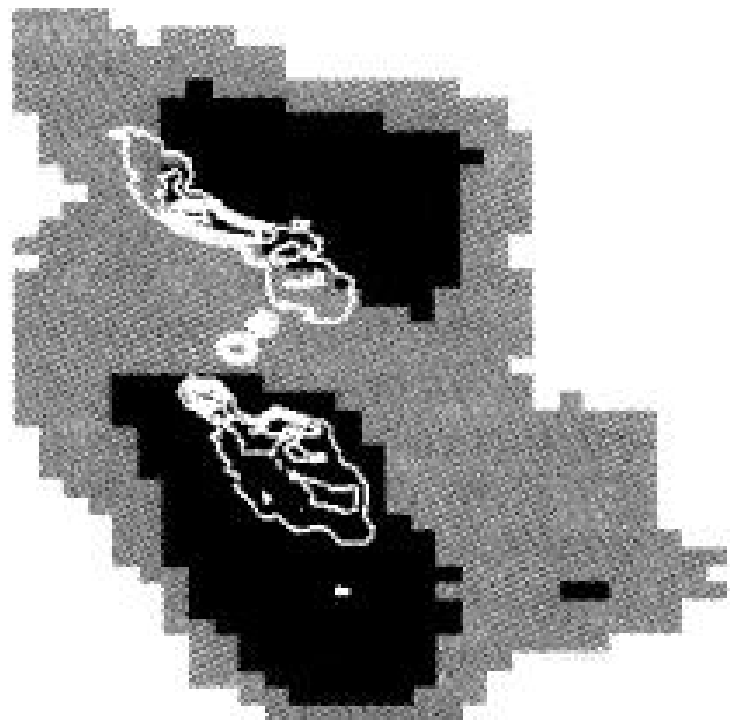}}}
\caption{ {\bf left} X-ray Chandra picture of the cooling flow at the center of
the galaxy cluster Abell 1795 (Fabian et al. 2001). The filament is about 
40kpc in length.
{\bf right} The cooling flow interacts at its base with the radio lobes, here represented
in contours, of the central radio source; the grey-scale indicates the presence
of blue-color regions, here in two patches at about 6 kpc from the nucleus,
just beyond the radio lobes (from Fabian 1994).}
\label{cooling}
\end{figure}

Another fueling mechanism could be the cooling flows that are expected to be 
infalling in the center of rich clusters. Indeed, it is frequent that AGN are 
present in the cluster central galaxies. They are frequenty powerful radio-sources,
 and the radio-lobes might be a heating/regulating process of the cooling gas 
(Fabian et al. 2001). The example of Abell 1795 in figure 
\ref{cooling} is illustrative of the phenomenon.

In summary, if the environment influence is evident for
powerful QSOs, Radio-galaxies and BlLacs, it is not so
significant for Seyferts.

\section{Galaxy Encounters and Binary BH Formation}

Given the large frequency of galaxy encounters
and mergers, if there is a massive black hole in nearly
every galaxy, the formation of a binary
black hole should be a common phenomenon.
The successive physical processes able to brake
the two black holes in their relative orbit have been
considered by Begelman et al (1980).

\subsection{Physical Processes Involved}

Each black hole sinks first toward the merger remnant center
through dynamical friction onto stars. A binary is formed;
but the life-time of such a binary can be much larger
than a Hubble time, if there is not enough stars to
replenish the loss cone, where stars are able to interact
with the binary.
Once a loss cone is created, it is replenished
only through the 2-body relaxation between stars,
and this can be very long (see section 1).
Modelising the merger remnant as an elliptical,
with a core of radius $r_c$ and mass $M_c$
(and corresponding velocity $V_c$), the radius
where loss cone effects are significant is:
$r_{lc}/r_c=(M_{bh}/M_c)^{3/4}$.
The various time-scales involved, and
corresponding characteristic scales are defined
by the following steps:
\begin{itemize}
\item the dynamical friction on stars, in less than a
galactic dynamical time,
$$
t_{df} \sim  (V_c/300{\rm km/s}) (r_c/100{\rm pc})^2 (10^8/M_{bh}) \, {\rm Myr}
$$
\item when the separation of the binary shrinks to
a value $r_b = r_c (M_{bh}/M_c)^{1/3}$, the two black holes become
bound together
\item the binary hardens, with $r_h \propto (r/r_b)^{3/2}$
\item  but when $r < r_{lc} = (M_{bh}/M_c)^{3/4} r_c$,
 the stars available for the binary to interact with, are depleted through
the loss cone effect, and replenished only by
2-body relaxation
\item gas infall can reduce the binary life-time
(whether the gas is flung out, or accreted, there is
a contraction of the binary) in $t_{gas}$
\item gravitational radiation shrinks the orbit on
$t_{GR} \sim 0.3 {\rm Myr} (10^8/M_{bh})^3 (r/0.003{\rm pc})^4$
if the two black holes have comparable masses
\end{itemize}

\subsection{Life-time of the Binary}

If the binary life-time is too long, another merger with
another galaxy will bring a third black-hole. Since a three-body
system is unstable, one of the three black-holes will be ejected
by the gravitational slingshot effect.

Since the life-time of the binary is not short,
there should be observable
manifestations of massive black hole binaries.
One of the best tracer is to detect the periodicity
of the keplerian motion, with
the period  P$\sim$ 1.6yr r$_{16}^{3/2}$ M$_8^{-1/2}$.
This is the case for the AGN
OJ 287 where eclipses have been monitored for a century
(Takalo 1994, Lehto \& Valtonen 1996, Pietil\"a 1998).
Also, if the black holes are rotating, and their
spins have misaligned axes, they precess around
the orbital one.
Plasma beams (aligned to the hole axis) precess,
and curved jets should be observed,
with periods between 10$^3$ to 10$^7$ yr.
This is frequently the case in
radio structures observed with VLA and VLBI,
modified by Doppler boosting, and light travel time
(cf 3C 273, NGC 6251, 1928+738, Kaastra \& Roos 1992;
Roos et al 1993).
Finally, pairs of radio galaxies have been observed during
their merger with four radio jets (3C75, Owen et al 1985).

Numerical simulations have brought more precision in
the determination of the life-time of the binary,
although numerical artifacts have given rise to debates.
Ebisuzaki et al (1991) claimed that the life-time of
the binary should be much shorter if its orbit is excentric,
since then the binary can interact with more stars and
release the loss cone problem. The first numerical
simulations tended to show that orbit
excentricity should grow quickly through
dynamical friction (Fukushige et al 1992).
Mikkola \& Valtonen (1992) and others found
that the excentricity in fact grows only very slowly.

To summarize the conclusions of several numerical
computations, there is finally
 little dependence on excentricity $e$, only in rare
cases, when $e$ is large from the beginning
(Quinlan 1996).
Eventually, the wandering of the binary helps the merging
of the two black holes (Quinlan \& Hernquist 1997).
The ejection out of the core of stars interacting with the binary
weakens the stellar cusp, while the
binary hardens. This may help to explain the
surprisingly weak stellar cusps in the center of giant ellipticals
observed recently with HST.
Observations show that bright elliptical
galaxies have weak cusps, while faint
galaxies have strong cusps, with a power law slope
of density versus radius of up to 2.
A way to weaken the cusps is a sinking black hole
(Nakano \& Makino 1999), or binary blaxk holes
(Milosavljevic \& Merritt, 2001) and this could be the
case for giant galaxies that have experienced
many mergers in their life.

\subsection{Conclusions on BH Binaries}

Binary black holes form efficiently during galaxy mergers, due to 
dynamical friction, in time-scales of 10-100 Myr.
The life-time of the binary is quite variable, 
and can be significantly shortened by gas accretion.
The slingshot effect could reduce the number of massive black holes,
but appears unlikely.

The rate of black holes merging is important to estimate
the possible detection rate of gravitational waves: 
if only bright AGN are taken into account, this rate is below 0.1 per year.
however, with a massive black hole in every galaxy with spheroid,
this rate reaches a few per year.
Massive black holes explain the formation of cusps in the stellar
density in the center of galaxies, while binary black holes are needed to 
flatten the cusp slopes.

\section{Cosmic Density of BH and Evolution with Redshift}

\subsection{Comparison with the Background Radiations}

An important observational result, from kinematical studies with high
 spatial resolution in nearby galaxies has been the existence of a black hole 
in each bulge/spheroid galaxy, with a proportionality factor between the mass 
of the BH and that of the bulge:
M$_{BH}$ = 0.002 M$_b$  (Magorrian et al 98) or an even better correlation with
the central stellar velocity dispersion (Ferrarese \& Merritt 2000,
Gebhardt et al. 2000). From this it is possible to deduce the 
density of black hole in the universe (given the baryonic density of the Universe).
 If we assume that this amount of mass of black hole has radiated while growing 
(i.e. masses of black holes are not above the Hills's limit), then it is interesting
 to compare the derived radiation density expected in the Universe from the 
formation of black holes. This comparison with the cosmic background radiations (e.g.
Haehnelt et al. 1998), assuming black holes radiate at Eddington luminosities,
 reveals that essentially the cosmic infra-red radiation is compatible with what
 expected from the present black hole density, assuming that a fraction of 30\% 
of it comes from AGN and the rest from starbursts. The optical cosmic background
 radiation is observed much lower, and suffers certainly from high obscuration.

\subsection{Evolution of the BH Mass in Active Galaxies}

\begin{figure}
{\centering \leavevmode
\epsfxsize=.40\textwidth \rotatebox{-90}{\epsfbox{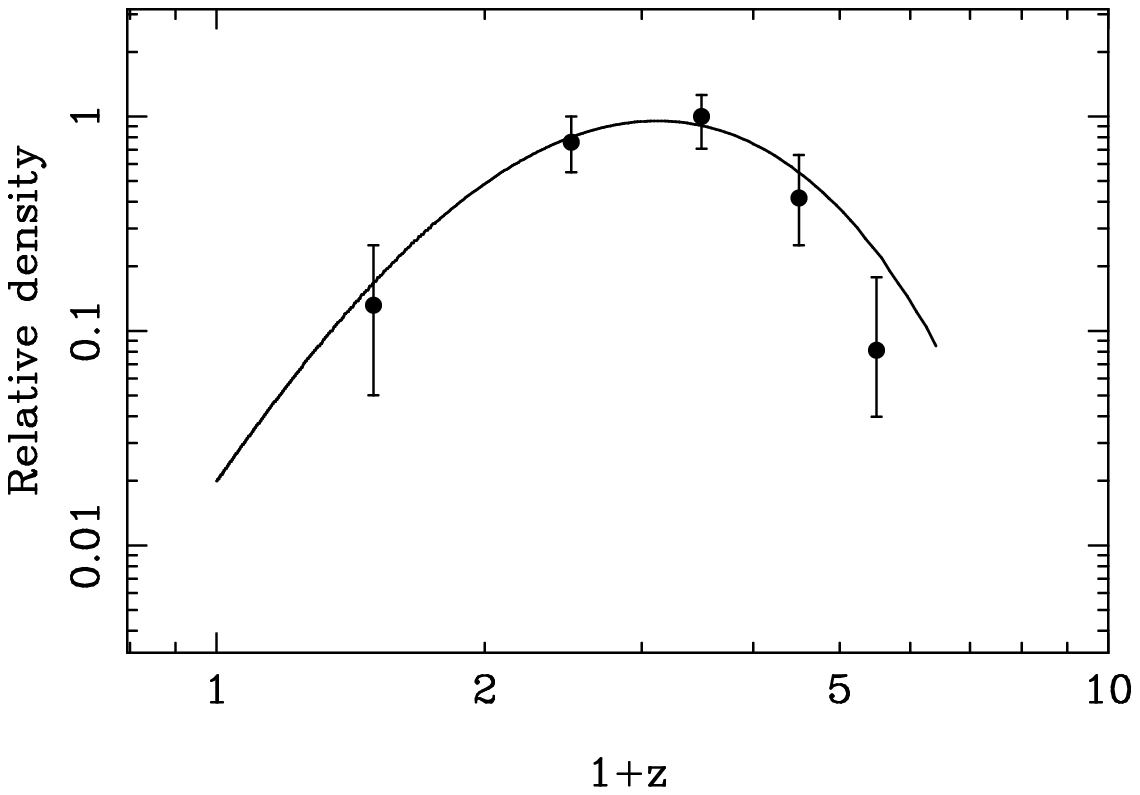}}\hfil
\epsfxsize=.40\textwidth \rotatebox{-90}{\epsfbox{combesf-f10b.ps}}}
\caption{ {\bf left} Space density as a function
of redshift, normalised to $z=2-3$, for the Parkes flat-spectrum
radio-loud quasars with P$_{11} >$ 7.2 10$^{26}$ W Hz$^{-1}$ sr$^{-1}$.
The optically-selected quasars follow the same curve (from Shaver
et al. 1996).
{\bf right} Cosmic history of star formation, for comparison.
The various data points, coming from different
surveys, give the universal metal ejection rate,
or the star formation rate SFR (left-scale),
as a function of redshift $z$. }
\label{shaver}
\end{figure}

Another observational data, to be taken into account to reproduce the formation 
of massive black holes, is the redshift evolution of the density of quasars, and 
the derived black hole masses. It is well known that the optically-selected quasars,
 but also the radio-loud quasars were more abundant in the past (Shaver et al 1996),
 and their evolution curve follows that of the density of starburst, or the star 
formation history (cf fig \ref{shaver}b).
Given the luminosity function of AGN (Boyle et al 1991), and assuming they are 
radiating at Eddington luminosity, a distribution of black hole mass can be 
derived. This shows that active galaxies in the past had in average larger 
black hole masses (Haehnelt \& Rees 1993).  

Of course, the mass of a given black hole only increases with time, but those
 which went through their active AGN phase in the past were of larger mass, 
because of a more abundant fueling, possibly due to the gas richness, and the 
shortest dynamical time in denser galactic systems.
At the present time, only low-mass spheroids (with low-mass BH) have sufficient
gas in their environment to become active. The large-mass black holes,
in big early-types objects like
massive ellipticals, are starving.

\section{Modelisation}

Semi-analytic models, based on the Press-Schechter formalism, and a CDM
hierarchical scenario of galaxy formation (Kauffmann \& Haehnelt 2000), can
reproduce rather well the essential observations: the proportionality relation between
the bulge and black hole mass in every galaxy, the amount of energy radiated
over the Hubble time due to accretion onto massive black holes, the past
evolution of AGN activity. The assumptions are that the black holes
grow through galaxy merging, both because of the merger of the
pre-existing black holes, and due to the infall of gas to the center in
the merging, that can fuel the merged BH. It is also assumed that
the cold gas in galaxies decrease with time; this implies that the
fueling will also decrease with time, accounting for the observed decline of AGN
activity. Finally, the gas accretion time-scale is proportional
to the dynamical time-scale, which is shorter at high redshift. The quasars
convert mass to energy at a fixed efficiency, and cannot radiate
more than the Eddington limit.

The results of such simulations are a strong decrease of the gas fraction in
galaxies, from 75\% at z=3 to 10\% at z=0, corresponding to the gas density
decrease observed in the damped Lyman alpha systems
(e.g. Storrie-Lombardi \& Wolfe 2000);
 this implies that the
black holes in the smallest ellipticals that formed at high z are relatively
more massive, since there was more gas at this epoch. Ellipticals forming
today have smaller black holes. Also the brightness of AGN for a given
galaxy was relatively higher in the past. The rapid decline of quasars
is then due to several causes:

\begin{itemize}
\item a decrease in the merging rate (which is also the cause of the
decrease of the star formation rate)

\item the decrease of the gas mass fraction in galaxies, due
to the consumption by stars

\item the increase of the accretion time-scales (the dynamical processes are  slower,
for a given amount of fuel)
\end{itemize}

\noindent In these kind of models, it is natural to expect a ratio of proportionality
between bulge and black hole masses, since they both form from the same
mechanisms, the hierarchical merging, and the corresponding
dynamical gas concentration. It is interesting to note that the life-time
of the quasar phase, a few 10$^7$ yr is then derived.

\section{Conclusions}

The rapid decline of AGN from z=2 to z=0 is parallel to that
of the star formation history, and is likely to have the same 
causes.  Since black hole masses have accumulated in the center
of each galaxy at the present time, the main reason for the
lower activity is the shortage of fuel. AGN are accreting today
at a rate much lower than Eddington rate. The dominant active
nuclei are presently low-luminosity AGN (Seyferts), revived
by gas accretion from galaxy interactions, much less frequent
than in the past. Also the galaxies with higher gas fraction are
late-type galaxies, that have small-mass black holes in their
nuclei. 

Spheroids and massive black holes appear to be formed 
with the same mechanisms.
Mass concentration is enhanced by radial gas flows, 
driven from the galactic disks,
through density waves, such as bars and spirals. 
Internal gravity torques
and non-axisymmetries are also boosted by galaxy interactions
and mergers.  Gas fueling towards the center can trigger
nuclear starbursts, forming
dense compact nuclear stellar clusters, that in turn will fuel
a massive black hole.  The detailed processes regulating the
successive/alternate fueling of starbursts and BH are not yet 
completely elucidated, and might involve the stability of galactic 
disks, the depth of the central potential well.

The required fueling depends on the strength and luminosity of the AGN.
For Seyferts, only stars from a dense nuclear cluster are sufficient,
through tidal disruptions and stellar collisions.
For quasars, big starbursts are required, and the coeval compact
cluster just formed can provide the fuel through mass loss of young stars 
and supernovae.

The relations between BH and bulge masses, or with the central
velocity dispersion, are then naturally explained by the
complicity between starbursts and AGN.

\end{document}